\documentclass[aps,twocolumn]{revtex4}
\usepackage{amssymb}

\usepackage{amsmath}
\usepackage{graphicx}

\input{tcilatex}

\begin{document}
\title{Many-body effects on the electronic and optical properties of strained
semiconducting carbon nanotubes}
\author{Catalin D. Spataru and Fran\c{c}ois L\'{e}onard}
\affiliation{Sandia National Laboratories, Livermore, California 94551}

\begin{abstract}
We present many-body \textit{ab initio} calculations of the electronic and
optical properties of semiconducting zigzag carbon nanotubes under uniaxial
strain. The GW approach is utilized to obtain the quasiparticle bandgaps and
is combined with the Bethe-Salpeter equation to obtain the optical absorption
spectrum. We find that the dependence of the electronic bandgaps on strain is
more complex than previously predicted based on tight-binding models or
density-functional theory. In addition, we show that the exciton energy and
exciton binding energy depend significantly on strain, with variations of tens
of meVs per percent strain, but that despite these strong changes the
absorbance is found to be nearly independent of strain. Our results provide
new guidance for the understanding and design of optomechanical systems based
on carbon nanotubes.

\end{abstract}
\maketitle

\section{INTRODUCTION}

The electronic and optical properties of nanomaterials such as carbon
nanotubes (CNTs), graphene, and nanowires show unique behavior due to their
reduced dimensions. For example, the electronic properties of CNTs depend
strongly on their diameter\cite{leonard}, and many-body effects are known to
significantly increase CNT bandgaps compared to density functional theory
(DFT)\cite{Spataru1}. In addition, CNT optical properties are dominated by
excitons \cite{Ando1,Spataru1,Chang1,perebeinos,Mazumdar1,Heinz1} due to a
combination between weak electrostatic screening \cite{leonard} and enhanced
Coulomb effects in quasi-one-dimensional systems \cite{Ando1,Spataru1}.

While the electronic and optical properties of isolated CNTs are well
understood, external factors such as the dielectric
environment\cite{perebeinos,Ando2}, electrostatic
doping\cite{spataru2,spataru3}, and nanotube-nanotube
interactions\cite{rohlfing}, have recently been shown to modify CNT electronic
and optical properties. Another important external factor that impacts CNT
properties is strain. Indeed, it was recognized early on\cite{heyd,tombler}
that strain can significantly modify CNT electronic properties, and this has
been exploited to realize new types of nanoelectromechanical
devices\cite{zheng}. However, to date theoretical studies of the impact of
strain on CNT electronic properties have been mostly limited to tight-binding
models and DFT; given the importance of many-body effects in unstrained CNTs,
a question to address is the role of many-body effects in strained CNTs.

The role of many-body effects on the optical properties of strained CNTs has
received even less attention. Experimental reports of strain modulation of CNT
optical properties have recently emerged\cite{maki,huang,kaniber}, and
indicate that optical transition energies can shift by tens of meVs per
percent strain. However, interpretation of these results has relied on
non-interacting models developed for electronic transitions, which do not
capture excitonic effects that dominate the optical response in CNTs. Progress
in developing exciton-based models for the optical properties of strained CNTs
has focused on approaches relying on the tight-binding or the $\mathbf{k\cdot
p}$ methods \cite{yu1,yu,yu2,ando}. However, a full many-body \textit{ab
initio} calculation of CNT optical properties under strain is still missing.

In this paper, we present such calculations by combining the GW approach with
the Bethe-Salpeter (BSE) equation to study the electronic and optical
properties of strained semiconducting CNTs. We find that the dependence of the
electronic bandgaps on strain is more complex than previously predicted based
on tight-binding models or density-functional theory. In addition, we show
that the exciton energy and exciton binding energy depend significantly on
strain, with variations of tens of meVs per percent strain. Furthermore, the
absorbance is found to be nearly independent of strain as a consequence of the
increase in transition dipole matrix elements with increasing strain.

This paper is organized as follows. After this Introduction, section II
describes the methodology and results for the electronic properties of
strained CNTs. Section III discusses the methodology and results for the
optical spectra, exciton energies, and excition binding energies. A summary is
presented in Section IV.

\section{Impact of many-body effects on electronic properties}

We perform our \textit{ab initio} calculations on the semiconducting (11,0)
and (17,0) CNTs for uniaxial strains from 0\% to 5\%. We start by
investigating the ground-state properties (e.g. relaxed atomic structure,
electron density) within DFT. The DFT calculations are performed using the
Quantum Espresso package\cite{espresso} within the Local Density Approximation
(LDA), using \textit{ab initio} pseudopotentials in combination with a
plane-wave basis set with a kinetic energy cutoff of 60 Ryd, in a supercell
geometry with tube separation (center to center) of more than double the
nanotube diameter. The atomic structure is relaxed until forces are smaller
than 5 meV/\AA \ for both the strained and unstrained cases. In the unstrained
case the (11,0) and (17,0) CNTs have diameters of 8.6 and 13.2
\AA \ respectively. The strain is applied by stretching (w.r.t. to the
unstrained case) the nanotube unit cell lattice vector along the tube axis
followed by atomic relaxation. This leads to a decrease in nanotube diameter
and a Poisson ratio $\nu\approx0.15$.

\begin{figure}[h]
\centering 
\includegraphics[width=3in]{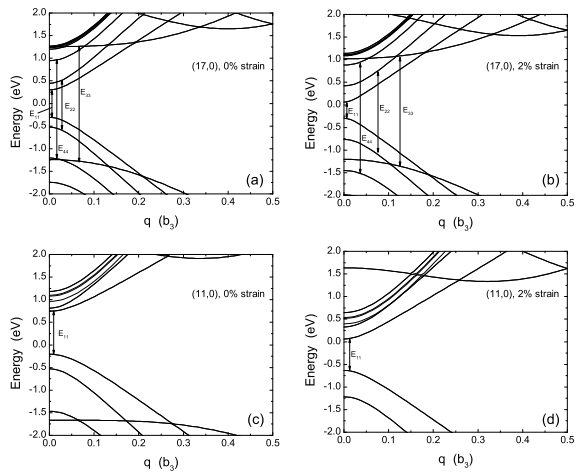}
\caption{DFT (LDA) bandstructure
of the (17,0) and (11,0) CNTs at 0\% and 2\% strain. The electronic bandgaps
and optical transitions studied in this paper are indicated with arrows. $b_{3}$ is the length of the Brillouin zone.}
\end{figure}

Figure 1 shows the DFT bandstructures at 0\% and 2\% strain, with the
electronic bandgaps and optical transitions studied in this paper labelled in
the figure. We note that for the unstrained (17,0) CNT the $E_{33}$ transition
has higher energy than the $E_{44}$ transition due to trigonal
warping\cite{trigonal}.Within LDA (Table 1), the (17,0) fundamental bandgap is
found to decrease with strain with a change $\Delta E_{g}^{11}$ of -125
meV/\%, in agreement with previous DFT calculations\cite{sreekala,valavala}.
The higher energy gaps show changes $\Delta E_{g}^{22}=+108$ meV/\%, $\Delta
E_{g}^{33}=-136$ meV/\%, and $\Delta E_{g}^{44}=+83$ meV/\%. These values can
be compared with those obtained from the simple tight-binding (TB) expression
for small strain\cite{yang} applied to zigzag CNTs%
\begin{equation}
\Delta E_{g}^{kk}=(-1)^{k}3\gamma(1+\nu)\sigma,
\end{equation}
where $\gamma$ is the tight-binding overlap integral and $\sigma$ is the
strain. Reasonable agreement with the LDA values can be obtained if one uses
$\gamma=3.3$ eV and $\nu=0.15$, giving $\Delta E_{g}/\sigma=$ $\pm$114 meV/\%.

While LDA and TB calculations agree to a large extent, an open question is
whether many-body effects can change the above picture. To address this
question, we performed quasiparticle calculations using the many-body GW
approach\cite{HL}. The electron self-energy $\Sigma=iGW$ is obtained within
the $G_{0}W_{0}$ approximation, i.e. using the LDA eigenvalues and
wavefunctions to construct the 1-particle Green's function $G$. The screened
Coulomb interaction $W$ is evaluated within the Random Phase Approximation and
extended at non-zero frequencies using the Plasmon-Pole approximation\cite{HL}%
. We consider empty states up to an energy cutoff of $\sim$60 eV, and use the
`static-remainder' technique\cite{statrem} to ensure convergence with respect
to the number of empty states. Convergence with respect to k-point sampling is
achieved with 128 k-points in the one-dimensional Brillouin zone. Also, the
Coulomb potential is truncated\cite{Sohrab,ApplPhysASpataru} in order to
prevent tube-tube interactions or periodic image effects due to the use of a
periodic supercell.

\begin{figure}[h]
\centering
\includegraphics[
width=3in
]%
{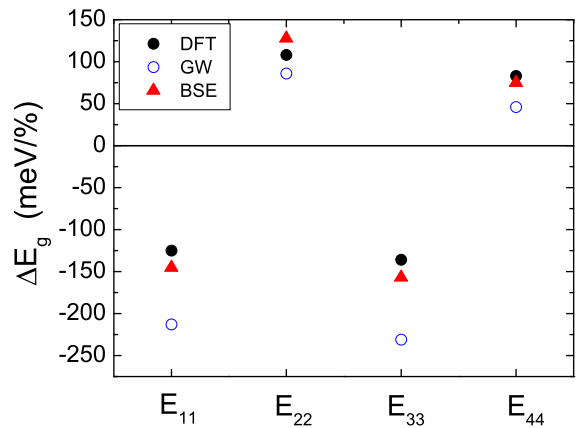}%
\caption{Strain dependence of the DFT, GW, and BSE gaps for the (17,0) CNT.}%
\end{figure}

Figure 2 and Table 1 show the calculated GW gaps as a function of strain for
the (17,0) CNT. The qualitative dependence on strain is similar to that
obtained within LDA, with the gaps increasing or decreasing with strain
depending on the band index. However, we find that strain effects are much
more complex within GW: for example $E_{11}$and $E_{33}$ decrease by 213
meV/\% and 231 meV/\% a much stronger change compared to LDA; on the other
hand $E_{22}$ and $E_{44}$ increase by 86 meV/\% and 46 meV/\%, much weaker
than LDA predicts.%

\begin{table}[tbp] \centering
\begin{tabular}
[c]{cccccc}\hline\hline
& Strain (\%) & E$_{11}$(eV) & E$_{22}$(eV) & E$_{33}$(eV) & E$_{44}%
$(eV)\\\hline
LDA & 0 & 0.606 & 0.968 & 2.495 & 2.153\\
& 2 & 0.356 & 1.184 & 2.223 & 2.329\\\hline
GW & 0 & 1.291 & 1.761 & 3.741 & 3.385\\
& 2 & 0.864 & 1.934 & 3.278 & 3.478\\\hline
BSE & 0 & 0.717 & 1.180 & 2.985 & 2.755\\
& 2 & 0.427 & 1.435 & 2.670 & 2.905\\\hline
$E_{b}$ & 0 & 0.574 & 0.581 & 0.756 & 0.630\\
& 2 & 0.437 & 0.499 & 0.608 & 0.573\\\hline\hline
\end{tabular}
\caption{Calculated transition energies for the (17,0) CNT using LDA, GW, and BSE. The exciton binding energy $E_b$ is
calculated as the difference between the transition energies from GW and BSE.}\label{TableKey}%
\end{table}%

The above results for the (17,0) CNT are limited to two strain values due to
the computational demands of the calculations. Tight-binding models predict a
linear dependence of the bandgaps on strain for relatively small strains with
changes in bandgaps independent of the tube diameter. To check whether this
trend holds when many-body effects are included, we performed GW calculations
for three different values of the strain for the $E_{11}$ gap of the (11,0)
CNT. As shown in Fig. 3 and Table 2, we obtain a linear dependence of the
bandgap on strain in agreement with the tight-binding prediction for small
strains and DFT. However, we find $\Delta E_{g}^{11}=-191$ meV/\% a value much
larger than the LDA value of $-127$ meV/\%; thus similar to the (17,0) CNT, we
find that many-body effects can significantly impact the electronic properties
of the strained (11,0) CNT.%
\begin{table}[tbp] \centering
\begin{tabular}
[c]{lllll}\hline\hline
Strain (\%) & LDA & GW & BSE & $E_{b}$\\\hline
0 & 0.950 & 1.922 & 1.062 & 0.860\\
2 & 0.694 & 1.540 & 0.775 & 0.764\\
5 & 0.319 & 0.837 & 0.348 & 0.488\\\hline\hline
\end{tabular}
\caption{Energies (in eV)  for the $E_{11}$ transition in the (11,0) CNT calculated using LDA, GW, and BSE. The exciton binding
energy $E_b$ is calculated as the difference between the GW and BSE energies.}\label{TableKey copy(1)}%
\end{table}%

\begin{figure}[h]%
\centering
\includegraphics[
width=3in]%
{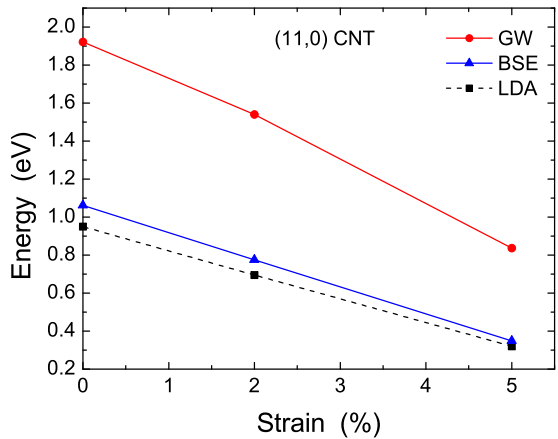}%
\caption{Strain dependence of the $E_{11}$transition for the (11,0) CNT.}%
\end{figure}

The differences between the GW and LDA results stem from the fact that with
reduction (augmentation) of the fundamental bandgap there is an increase
(decrease) in the dielectric screening $\varepsilon$ of the CNT. Consider the
case where the fundamental bandgap decreases with strain: we plot in Fig. 4
the dielectric screening $\varepsilon^{-1}\left(  q,\omega=0\right)  $ for the
(11,0) CNT for strains of 0\%, 2\%, and 5 \%. The increased screening affects
the screened Coulomb interaction $W=\varepsilon^{-1}v$ and hence the electron
self-energy $\Sigma=iGW$ present in the many-body calculations. More exactly,
the contribution $\Sigma_{g}$ of the electron self-energy to the quasiparticle
bandgap $E_{g}^{GW}=E_{g}^{LDA}-V_{g}^{xc}+\Sigma_{g}$, decreases appreciably
(same order of magnitude as the change in the LDA bandgap) upon strain:
$\delta\Sigma_{g}\equiv\Sigma_{g}(\sigma)-\Sigma_{g}(\sigma=0)<0$. This is a
many-body effect not captured by the LDA exchange-correlation Kohn-Sham
potential $V^{xc}$, and as expected we find $\delta V_{g}^{xc}/\delta
\Sigma_{g}\ll1$. Thus, the change in the fundamental quasiparticle bandgap
upon strain as obtained within GW is more pronounced than the one obtained
within a mean-field (LDA) theory, with appreciable contribution from
self-energy corrections: $\delta E_{g}^{GW}\approx\delta E_{g}^{LDA}%
+\delta\Sigma_{g}$. For $E_{11}$ and $E_{33}$ this leads to a larger decrease
in the bandgap compared to LDA since both $\delta E_{g}^{LDA}$ and
$\delta\Sigma_{g}$ are negative; in contrast, $\delta E_{g}^{LDA}$ is positive
for $E_{22}$ and $E_{44}$, and the still negative $\delta\Sigma_{g}$ leads to
a smaller increase of the bandgap with strain.

\begin{figure}[h]%
\centering
\includegraphics[width=3in
]%
{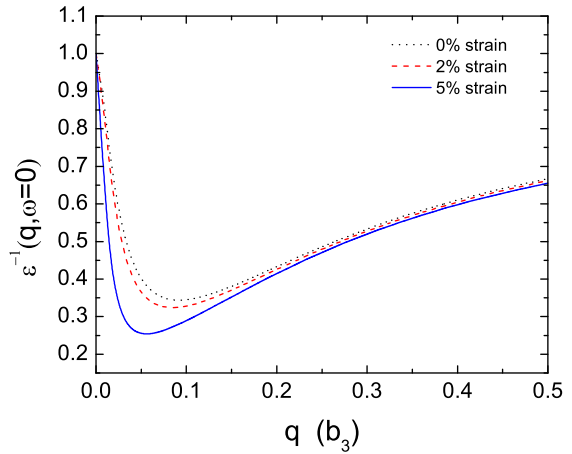}%
\caption{Strain dependence of the dielectric screening $\varepsilon^{-1}$ for
the (11,0) CNT. $b_{3}$ is the length of the unit cell.}%
\end{figure}

\section{Impact of many-body effects on optical properties}

We next turn to the optical properties. To calculate these, we start from the
GW results and couple them with the BSE. Both the BSE and GW calculations were
performed using the BerkeleyGW package\cite{BerkeleyGW}. We solve the BSE for
excitons within the static approximation for the dielectric screening and
within the Tamm-Dancoff approximation for excitons \cite{Tamm}. Having
obtained the excitonic properties one can then obtain the optical response of
CNTs using the standard approach \cite{Spataru1,BerkeleyGW}. The optical
bandgap is equal to the quasiparticle bandgap minus the binding energy of the
lowest bright exciton, a quantity which results from the overall attractive
electron-hole interaction between the (quasi)electron and the (quasi)hole
forming the exciton.

Figure 5 shows the optical absorbance for the (17,0) CNT calculated within GW
(no excitonic effects) and calculated within BSE (i.e. including excitonic
effects) for the two lowest optical transitions. Here the absorbance is
obtained from $A(\omega)\sim\omega\varepsilon_{2}\left(  \omega\right)  $
where $\varepsilon_{2}\left(  \omega\right)  $ is the imaginary part of
$\varepsilon$. The peaks in the figure for the BSE results indicate the lowest
energy bright exciton for light polarization parallel to the nanotube axis. At
zero strain (Fig. 5a), the $E_{11}$ and $E_{22}$ transitions show strong
many-body effects, with exciton binding energies of 574 meV and 581 meV.
Similar results are obtained for the $E_{33}$ and $E_{44}$ transitions (Table
1) with $E_{b}^{33}=756$ meV and $E_{b}^{33}=630$ meV.

Upon application of strain (Fig. 5b) one can see that the $E_{11}$exciton
energy $\Omega_{11}$decreases by 145 meV/\% while $\Omega_{22}$ increases by
128 meV/\%. Results for $\Omega_{33}$ and $\Omega_{44}$ (Table 1) give values
of -157 meV/\% and +75 meV/\%, respectively. Thus, the qualitative trends
observed from the GW calculations are maintained with the optical properties;
however, because the quasiparticle bandgap and the exciton energy have a
different dependence on strain, $dE_{g}^{GW}/d\sigma\neq d\Omega/d\sigma$, one
can deduce that the exciton binding energy $E_{b}=E_{g}^{GW}-\Omega$ depends
on strain. This can be seen in Fig. 5 and in Table 1 where all of the exciton
binding energies are decreased under strain by amounts ranging from 28 meV/\%
for the $E_{44}$ transition to 74 meV/\% for the $E_{33}$ transition. Much
like the changes in the quasiparticle gap, the decrease in binding energy also
stems from the change in dielectric screening upon applied strain. Indeed, the
attractive interaction between the electron and hole forming the exciton is
mediated by the screened Coulomb interaction $W=\varepsilon^{-1}v$, and
because $\varepsilon$ is always determined by the lowest energy electronic
bandgap, all of the optical transitions will be affected in the same way
leading to the common decrease in binding energy.

It should also be noted that the exciton oscillator strength shows very small
variation with strain. Since the oscillator strength is $\sim\Omega\mu_{a}%
^{2}/a$, with $\mu_{a}^{2}/a$ the squared exciton transition dipole matrix
element per unit tube length \cite{Spataru2}, the implication is that $\mu
_{a}^{2}/a$ strongly increases with increasing strain. Indeed, for the (17,0)
CNT, we find that $\mu_{a}^{2}/a$ increases from $\sim$3.9 a.u. at zero strain
to $\sim$6.8 a.u. at 2\% strain.

\begin{figure}[h]%
\centering
\includegraphics[width=3in]{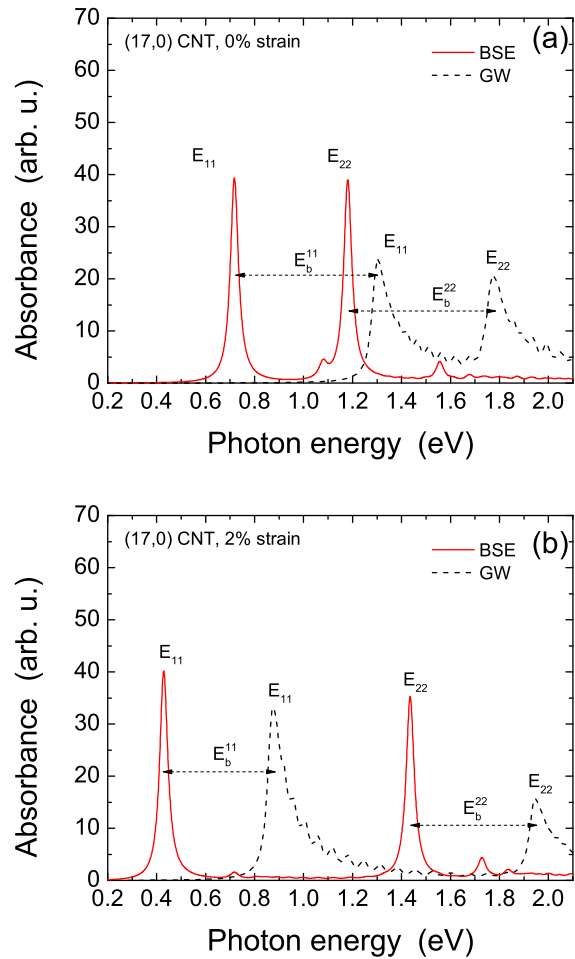}%
\caption{Optical absorption spectrum of the (17,0) CNT calculated without the
electron-hole interaction (GW) and with the electron-hole interaction (BSE).
Panel (a) is for the unstrained case and panel (b) is for 2\% strain.}%
\end{figure}

The optical results for the (17,0) CNT can be generalized to the (11,0) CNT as
well, at least for the lowest optical transition (Fig. 3 and Table 2). Indeed
we find $d\Omega^{11}/d\sigma=-142$ meV/\% and a reduction of the exciton
binding energy by severals tens of meV/\%. (At 2\% the reduction in $E_{b}$ is
about 3.5 times as large as that obtained using a tight-binding approach for
excitons\cite{yu} for the (11,0) CNT.) Furthermore, the exciton energy is
found to depend linearly on strain, and turns out to be relatively close to
the DFT result. As we discussed above, with decreasing bandgap the dielectric
screening gets enhanced, and thus the binding between electron and hole
decreases. This effect also explains why the change in optical gap $\Omega$
upon applied strain is similar to that obtained at the LDA level: it is due to
cancellation effects between quasiparticle self-energy corrections and
excitonic effects.

\section{Summary}

In summary, we performed many-body \textit{ab initio} calculations of the
electronic and optical properties of semiconducting zigzag CNTs under uniaxial
strain. We find that the fundamental electronic bandgap depends more strongly
on strain than previously predicted by non-interacting models. In addition, we
find that self-energy corrections generally decrease the bandgaps, which
enhances or reduces the impact of strain compared to DFT depending on which
transition is considered. Furthermore, the optical transitions are also found
to be affected by many-body effects. In particular, the exciton binding energy
decreases with increasing strain regardless of the transition, with variations
of several tens of meVs per percent strain. More generally, our results
indicate that quasiparticle and excitonic effects are strongly tied, and that
the interpretation of optomechanical experiments in CNTs requires a more
in-depth consideration of many-body effects. This is further supported by
other many-body calculations on strained bulk\cite{thulin, lambrecht} and
two-dimensional materials\cite{shi,liang} where material-specific and
dimensionality phenomena have been observed.

\textbf{ACKNOWLEDGEMENTS}

Work supported by the Laboratory Directed Research and Development program at
Sandia National Laboratories, a multiprogram laboratory managed and operated
by Sandia Corporation, a wholly owned subsidiary of Lockheed Martin
Corporation, for the United States Department of Energy's National Nuclear
Security Administration under Contract DE-AC04-94AL85000.

\end{document}